\documentclass[useAMS,usenatbib]{mn2e}

\usepackage{graphics}
\usepackage{graphicx}

\title[Spectroscopic Observations of Five Comets]
  {Spectroscopic Observations of New Oort Cloud Comet 2006 VZ13 and Four Other Comets}
\author[A.M. Gilbert et al.]
  {A.M.~Gilbert,$^1$\thanks{Corresponding author email: alyssa.gilbert@uwo.ca}
  P.A.~Wiegert,$^1$ E.~Unda-Sanzana,$^2$ O.~Vaduvescu$^{2,3}$\\
  $^1$Department of Physics \& Astronomy, The University of Western Ontario, London, Ontario, Canada N6A 3K7\\
  $^2$Instituto de Astronom\'{i}a, Universidad Cat\'{o}lica del Norte, Avenida Angamos 0610, Antofagasta, Chile\\
  $^3$Isaac Newton Group of Telescopes, Apartado de Correos 321, E-38700 Santa Cruz de la Palma, Canary Islands, Spain}
\date{Released 2002 Xxxxx XX}

\pagerange{\pageref{firstpage}--\pageref{lastpage}} \pubyear{2002}

\def\LaTeX{L\kern-.36em\raise.3ex\hbox{a}\kern-.15em
    T\kern-.1667em\lower.7ex\hbox{E}\kern-.125emX}

\begin{document}

\label{firstpage}

\maketitle

\begin{abstract}
Spectral data are presented for comets 2006~VZ13~(LINEAR), 2006~K4~(NEAT), 2006~OF2~(Broughton), 2P/Encke, and 93P/Lovas~I, obtained with the Cerro-Tololo Inter-American Observatory 1.5-m telescope in August 2007. Comet 2006~VZ13 is a new Oort cloud comet and shows strong lines of CN (3880~\AA{}), the Swan band sequence for C$_2$ (4740, 5160, and 5630~\AA{}), C$_3$ (4056~\AA{}), and other faint species. Lines are also identified in the spectra of the other comets. Flux measurements of the CN, C$_2(\Delta v=+1,0)$, and C$_3$ lines are recorded for each comet and production rates and ratios are derived. When considering the comets as a group, there is a correlation of C$_2$ and C$_3$ production with CN, but there is no conclusive evidence that the production rate ratios depend on heliocentric distance. The continuum is also measured, and the dust production and dust-to-gas ratios are calculated. There is a general trend, for the group of comets, between the dust-to-gas ratio and heliocentric distance, but it does not depend on dynamical age or class. Comet 2006~VZ13 is determined to be in the carbon-depleted (or Tempel~1 type) class.

\end{abstract}


\section{INTRODUCTION}
The icy nature of comets indicate they have been preserved at cold temperatures since the early stages of solar system formation. Consequently, they are commonly considered to be among the most primitive objects in the solar system. Determining their physical and chemical properties, and how they evolve, is important to our understanding of the formation of planetary systems, both our own and in general.

Several surveys have attempted to study and classify the chemical composition and evolution of comets. \citet{A1995} conducted a photometric survey of 85 different comets over almost 20 years. Their findings indicated two major classes of comets: those that are carbon-depleted and those that are not. They found nearly all members of the carbon-depleted class are Jupiter family comets (JFCs), although not all JFCs are carbon-depleted. They also reported little variation of relative production rates with heliocentric distance or apparition; however, they noted a correlation between the dust-to-gas ratio and perihelion distance. 

Three major spectroscopic surveys have also been conducted: \citet{NS1989} reported spectrophotometry of 25 comets; \citet{C1992} derived production rates for 17 faint comets; and \citet{FH1996} surveyed the spectra of 39 comets into infrared wavelengths. The three surveys found slightly different results. \citet{NS1989} reported a correlation between CN and dust, and that the C$_2$/CN production ratio changed with heliocentric distance. \citet{C1992}, however, found the gas production ratios remained constant with activity level and heliocentric distance, except for NH$_2$/CN. \citet{FH1996} concluded that most comets have roughly the same production rate ratios to within a factor of 2 or 3, although 10~per~cent of comets could be considered outliers.

More recently, \citet{F2009} presented a spectroscopic survey of 92 comets over approximately 19 years. They report four taxanomic classes of comets: typical, Tempel~1 type, Giacobini-Zinner type, and the unusual object Yanaka (1988r). The typical comets have typical ratios of C$_2$, NH$_2$, and CN with respect to water, while Tempel-1 types have deficiencies in C$_2$ but normal NH$_2$ abundances. Giacobini-Zinner comets have low C$_2$ and NH$_2$ ratios with respect to water, while Yanaka has no detectable C$_2$ or CN emission, but normal NH$_2$ abundances. They conclude that the Halley family of comets (originating in the Saturn and Uranus region, but were scattered to the Oort cloud) shows no C$_2$ depletion, while objects originating in the Neptune region show a mixture of typical and C$_2$ depleted objects. Comets originating in the classical Kuiper belt form the C$_2$ depleted group.

Comet 2006~VZ13~(LINEAR) was discovered in November 2006 \citep{S2006}, and is a new Oort cloud comet which passed perihelion on 10 August 2007. This presented a unique opportunity to observe a pristine comet as it passed perihelion. In addition, four other comets were observed with the same instrument and under the same observing conditions: 2006~K4~(NEAT), 2006~OF2~(Broughton), 93P/Lovas~I, and 2P/Encke. These comets represent a broad range of dynamical class, age, brightness and heliocentric distance. This allows for a comparison of production rates and ratios between comets of different classes and ages.
\begin{table*}
 \centering
 \begin{minipage}{150mm}
  \caption{Orbital elements of the comets.\setcounter{footnote}{\value{footnote}}\protect\footnotemark}\label{tab:orbits}
  \begin{tabular}{@{}lcccccccc@{}}
  \hline
Comet & $a$ (au) & $e$ & $i$ ($^o$) & $q$ (au) & $Q$ (au) & $T_J$ & Perihelion Date (UT) & Comet Class\setcounter{footnote}{\value{footnote}}\protect\footnotemark \\ \hline
2P/Encke	&	2.217	&	0.847	&	11.766	&	0.339	&	4.095	& 3.026 & 2007 Apr 19 & NEO\\ 
2006 K4		&  1805.756  &	0.998	&	111.333	&	3.189	&	3608.323 &	-0.802 & 2007 Nov 29 & EOC\\ 
2006 OF2	&  -3367.756  &	1.001	&	30.171	&	2.431	&	n/a 	& 	n/a & 2008 Sep 15 & NOC\\
2006 VZ13	&  -4083.355  &	1.000	&	134.793 &	1.015   &	n/a	&	n/a & 2007 Aug 10 & NOC\\ 
93P/Lovas~1	&  4.391	&	0.612	&	12.218	&	1.705	&	7.078	& 2.605 & 2007 Dec 17 & JFC\\ \hline
\end{tabular}
\footnotetext{$^1$ All values are from the JPL Small-Body Database Browser; $a$ is the semi-major axis, $e$ is the eccentricity, $i$ is the inclination, $q$ is the perihelion distance, $Q$ is the aphelion distance, and $T_J$ is the Jupiter Tisserand parameter (given by $T_J=a_J/a +2\cos i\sqrt{a(1-e^2)/a_J}$).}
\footnotetext{$^2$ NEO = near earth object; EOC = evolved Oort cloud comet; NOC = new Oort cloud comet; JFC = Jupiter family comet.}
\end{minipage}
\end{table*}

\setcounter{footnote}{0}
\begin{table*}
 \centering
 \begin{minipage}{105mm}
  \caption{CTIO observational parameters.\setcounter{footnote}{\value{footnote}}\protect\footnotemark}\label{tab:obs}
  \begin{tabular}{@{}lccccc@{}}
  \hline
Comet & Date (UT) & $\Delta$ (au) & $r_H$ (au) & $V$ (mag) & Exp. Time (s) \\ \hline
2P		&	07/08/07 00:16	&	0.9400	&	1.9114	&	15.19	&	3600\\
		&	07/08/07 23:56	&	0.9550	&	1.9221	&	15.30	&	3650\\
		&	13/08/07 01:30	&	1.0381	&	1.9781	&	15.92	&	5400\\
		&	14/08/07 01:14	&	1.0551	&	1.9890	&	16.04	&	5400\\ 
2006 K4		&	07/08/07 01:24	&	2.6270	&	3.3713	&	15.75	&	3600\\
	&	08/08/07 02:08	&	2.6364	&	3.3681	&	15.75	&	3600\\
	&	13/08/07 03:33	&	2.6869	&	3.3529	&	15.78	&	5400\\
	&	14/08/07 03:04	&	2.6977	&	3.3500	&	15.78	&	5400\\
	&	15/08/07 02:33	&	2.7084	&	3.3472	&	15.79	&	5400\\ 
2006 OF2	&	07/08/07 02:53	&	3.7934	&	4.7860	&	16.33	&	3600\\
	&	08/08/07 03:19	&	3.7825	&	4.7782	&	16.32	&	3600\\
	&	13/08/07 06:32	&	3.7316	&	4.7382	&	16.26	&	5400\\
	&	14/08/07 06:17	&	3.7226	&	4.7304	&	16.25	&	5400\\
	&	15/08/07 05:06	&	3.7143	&	4.7229	&	16.24	&	5400\\ 
2006 VZ13	&	06/08/07 23:14	&	1.0207	&	1.0175	&	13.26	&	1800\\
	&	07/08/07 23:14	&	1.0482	&	1.0165	&	13.32	&	1800\\
	&	12/08/07 23:22	&	1.1866	&	1.0159	&	13.58	&	3600\\
	&	13/08/07 23:19	&	1.2142	&	1.0166	&	13.64	&	3600\\
93P		&	08/08/07 06:03	&	1.4603	&	2.1451	&	16.61	&	7200\\
		&	15/08/07 07:36	&	1.3617	&	2.1059	&	16.38	&	9600\\ \hline
\end{tabular}
\footnotetext{$^1$ The date indicates the start time of the first observation of the comet; $\Delta$ is the geocentric distance, $r_H$ is the heliocentric distance, and $V$ is the total magnitude. The exposure time is the total of all observations on the given night.}
\end{minipage}
\end{table*}
In this paper, spectroscopic observations are presented of five comets, obtained in August 2007. The production rates of CN, C$_2$, and C$_3$ are calculated, and the production ratios with respect to CN are derived. In addition, dust production rates and dust-to-gas ratios are calculated.

\section{OBSERVATIONS}
Observations of 2006~VZ13, 2006~K4, 2006~OF2, 2P, and 93P were obtained during 2007 August 6--15 (UT). Table~\ref{tab:orbits} lists the orbital parameters for each comet, and Table~\ref{tab:obs} summarises the observations. All observations were taken under photometric conditions, except for the second half of the night of August 14, which was partly cloudy.

All observations were acquired with the Cerro-Tololo Inter-American Observatory (CTIO) 1.5-m telescope and R--C spectrograph. The spectrograph had a Loral 1K $1200\times800$ CCD with 15-$\mu$m pixels. A Bausch \& Lomb reflection grating was used with 300 lines mm$^{-1}$, a resolution of 8.6~\AA{}, and a coverage of 3450~\AA{}. A 2-arcsec slit was used for all observations. Wavelength scales were approximately 3~\AA{} pixel$^{-1}$ from 3500 to 7000~\AA{}. The CCD spatial scale was 1.13 arcsec pixel$^{-1}$, with the length of the slit spanning 460 arcsec. 

Each comet nucleus was centred on the slit for the duration of the exposure. All of the comets, with the exception of 2006 VZ13, only covered a small area on the slit (on the order of the width of the slit). This is due to the objects being at large distances and having relatively small comae. Since 2006 VZ13 covered the entire slit, this could have introduced errors in guiding and pointing of the telescope, potentially causing a misalignment of the nucleus on the slit. Great care was taken during the observations to track the object using the coordinates from the JPL database in order to minimize these errors.

Bias and projector flats were obtained at the beginning of each night, and HeAr comparison lamp spectra were recorded before each new object. Images of the twilight sky were also obtained, and were used to partially correct for solar reflection. Standard star Feige~110 was observed to flux calibrate the comet spectra.

\section{DATA REDUCTION}
The data was reduced using IRAF \citep{T1993}. For each observation date, the projector flats were examined to determine portions of the chip that were not illuminated. These sections were trimmed from all images. The bias frames were combined and applied to each twilight sky flat, projector flat, lamp spectrum and object spectrum. The projector flat frames were then combined and applied to the sky flats and object spectra.

One-dimensional spectra of each sky flat and object were extracted using IRAF's APALL tool, which traced the centre of the profile, corrected cosmic-ray hits, and subtracted the background sky near the edges of the slit (with the exception of 2006 VZ13, where the background sky correction was taken from another image). Each spectrum was wavelength calibrated using a HeAr lamp spectrum. Spectra obtained of the same object on the same night were co-added to increase the signal-to-noise ratio.

The combined sky flat image was subtracted from the cometary spectra to partially correct for solar reflection. This continuum correction was not perfect and, in general, left 10--20~per~cent of the continuum. To compensate for the difference, the continuum of each comet was fit again with a cubic spline function and subtracted. 

The comet spectra were flux calibrated using spectra of the standard star, Feige 110. The star was observed several times per night at different airmasses. Using these observations, the effective airmass and system sensitivity function were calculated, which was then used to calibrate the comet spectra. 

\begin{figure*}
\includegraphics[width=5in]{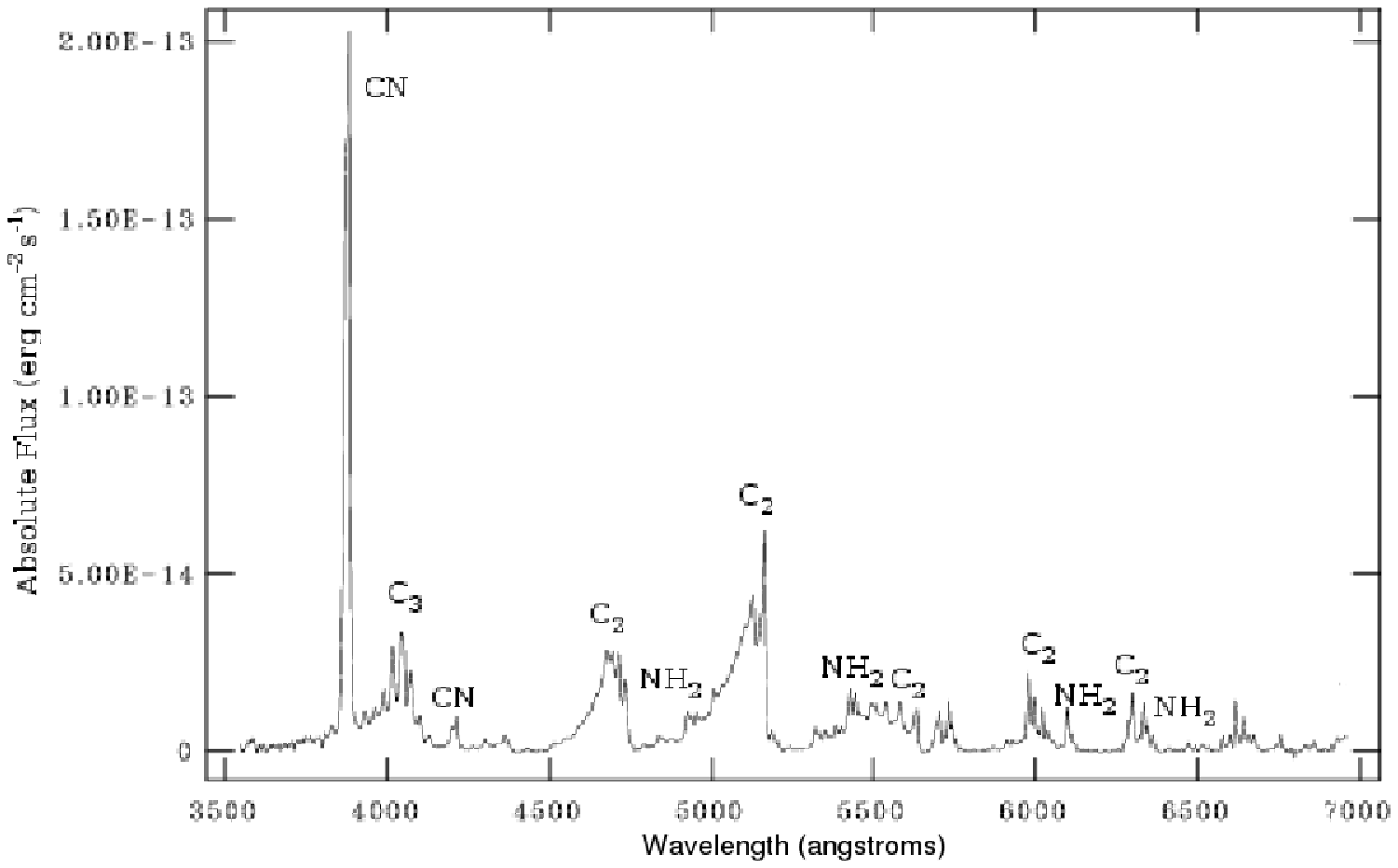}
  \caption{Continuum-subtracted, absolute-flux-calibrated spectrum of comet 2006~VZ13, obtained 2007 August 13. Major emission lines are labelled.}
\label{fig:ex}
\end{figure*}

\begin{figure*}
\includegraphics[width=5in]{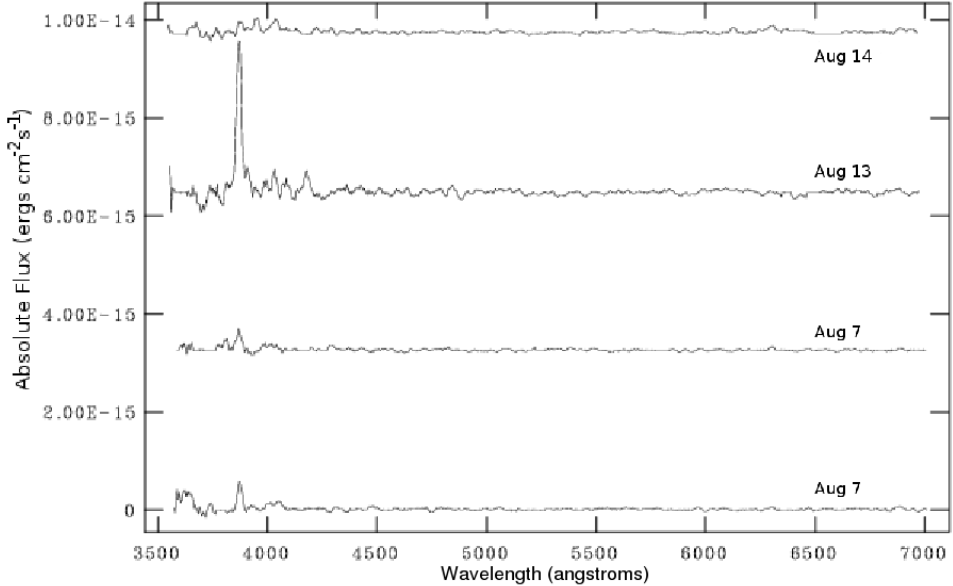}
  \caption{Four continuum-subtracted optical spectra of comet 2P/Encke, obtained in August 2007. Spectra are offset by a factor of $3.25\times10^{-15}$ for clarity. Note the strong increase of the CN line on August 13.}
\label{fig:2P_labels}
\end{figure*}

\begin{figure*}
\includegraphics[width=5in]{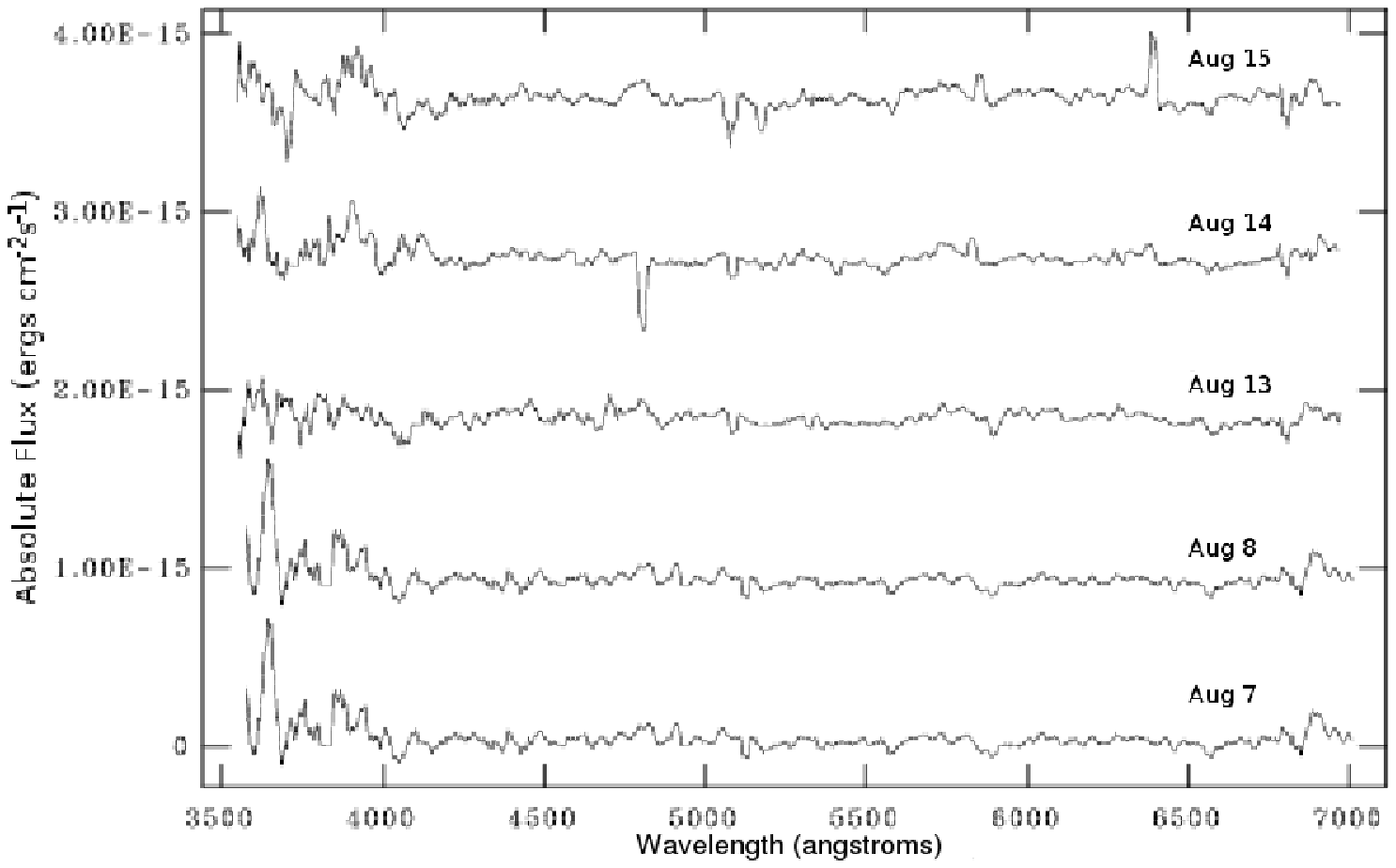}
  \caption{Five continuum-subtracted optical spectra of comet 2006~K4, obtained in August 2007. Spectra are offset by a factor of $9.0\times10^{-16}$ for clarity.}
\label{fig:2006K4_labels}
\end{figure*}

\begin{figure*}
\includegraphics[width=5in]{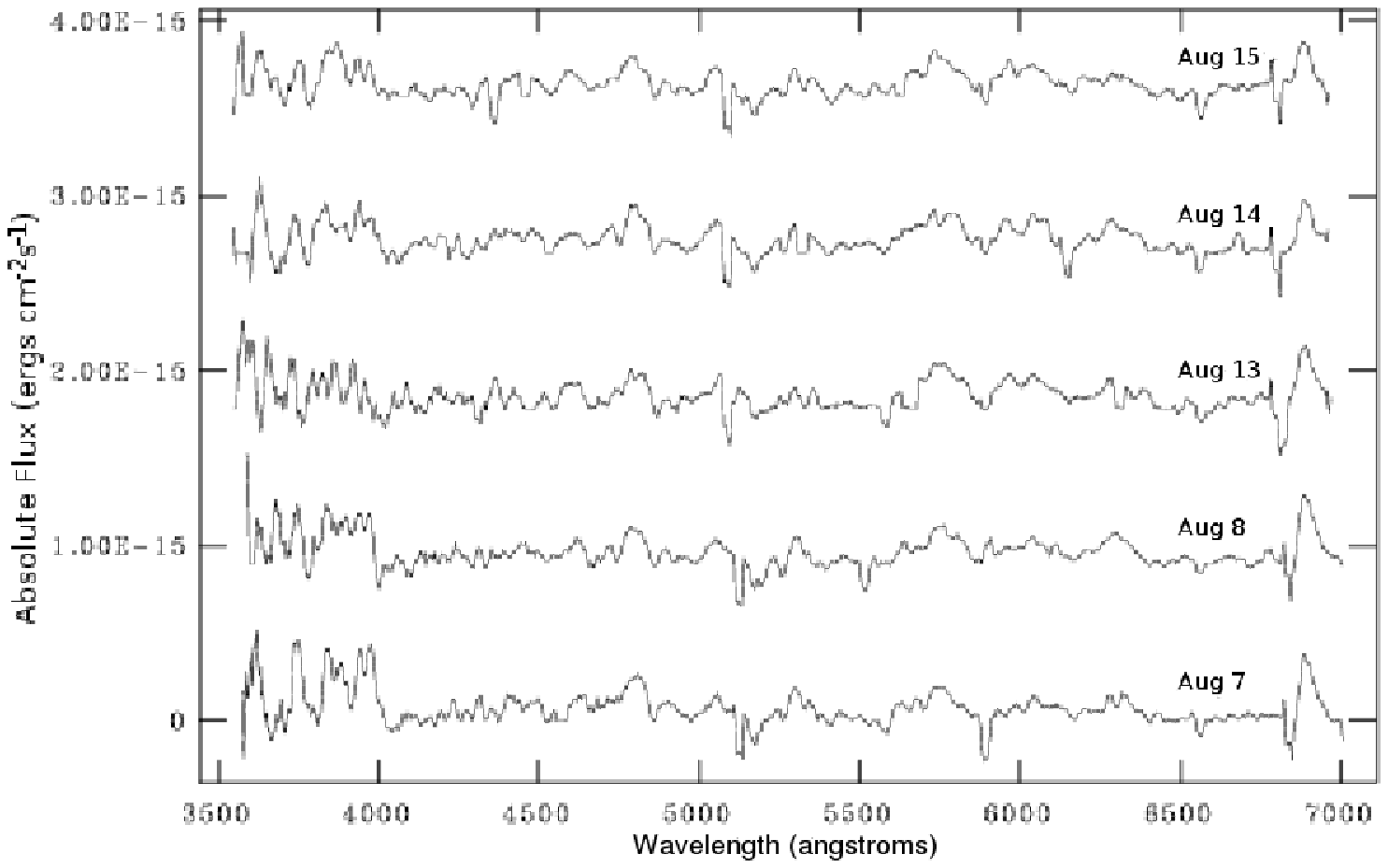}
  \caption{Five continuum-subtracted optical spectra of comet 2006~OF2, obtained in August 2007. Spectra are offset by a factor of $9.0\times10^{-16}$ for clarity.}
\label{fig:2006OF2_labels}
\end{figure*}

\begin{figure*}
\includegraphics[width=5in]{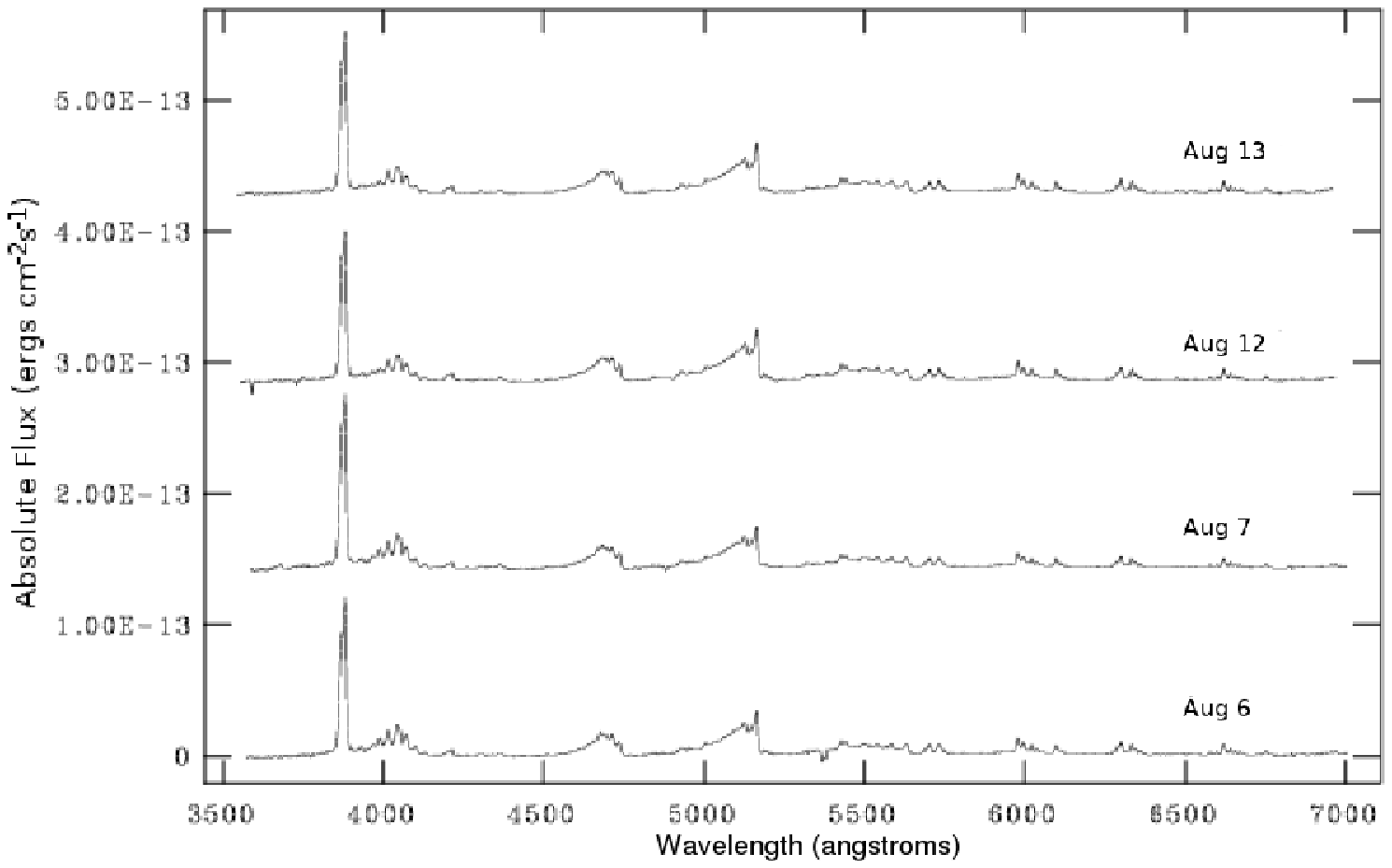}
  \caption{Four continuum-subtracted optical spectra of comet 2006~VZ13, obtained in August 2007. Spectra are offset by a factor of $1.45\times10^{-13}$ for clarity.}
\label{fig:VZ13_labels}
\end{figure*}

\begin{figure*}
\includegraphics[width=5in]{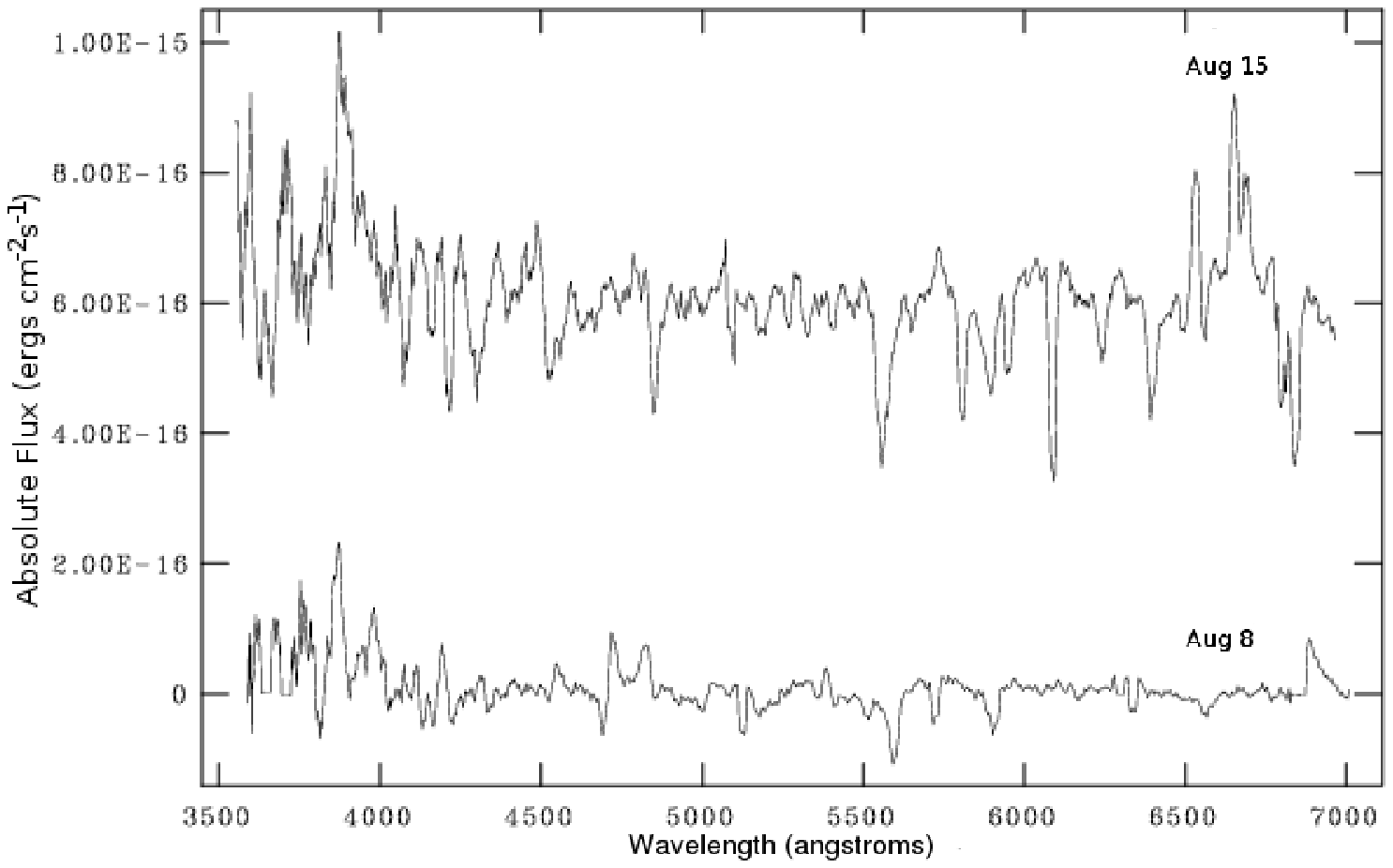}
  \caption{Two continuum-subtracted optical spectra of comet 93P, obtained in August 2007. Spectra are offset by a factor of $6.0\times10^{-16}$ for clarity.}
\label{fig:93P_labels}
\end{figure*}

Fig.~\ref{fig:ex} shows the 2007 August 13 spectrum of 2006~VZ13 with the emission lines labelled. Figs.~\ref{fig:2P_labels}--\ref{fig:93P_labels} show the final one-dimensional, absolute-flux-calibrated spectra of each comet. All spectra are binned by nine pixels to improve the signal-to-noise ratio (excluding 2006~VZ13).


\section{ANALYSIS AND RESULTS}
The absolute flux is measured above the continuum of the CN, C$_3$ and C$_2$ emission features for each comet (see Table~\ref{tab:bands} for parameters). The continuum flux is also measured to calculate the dust production rate. The measured fluxes are listed in Table~\ref{tab:fluxes}.

\setcounter{footnote}{0}
\begin{table*}
 \centering
 \begin{minipage}{115mm}
  \caption{Molecular parameters.\setcounter{footnote}{\value{footnote}}\protect\footnotemark}\label{tab:bands}
  \begin{tabular}{@{}lcccc@{}}
  \hline
Molecule & Wavelength Range (\AA{}) & $g$ (erg s$^{-1}$ mol$^{-1}$) & $l_p$ (km) & $l_d$ (km) \\ \hline
CN ($\Delta v=0$)	&	$3840-3900$	&	$2.4-5.0\times10^{-13}$	&	$1.3\times10^4$	&	$2.1\times10^5$ \\
C$_3$ 			&	$3950-4100$	&	$1.0\times10^{-12}$	&	$2.8\times10^3$	&	$2.7\times10^4$ \\
C$_2$ ($\Delta v=+1$)	& 	$4550-4750$ 	&	$4.5\times10^{-13}$	&	$2.2\times10^4$	&	$6.6\times10^4$ \\
C$_2$ ($\Delta v=0$)	& 	$5000-5200$	&	$4.5\times10^{-13}$	&	$2.2\times10^4$	&	$6.6\times10^4$ \\
Continuum		&	$6230-6270$	&				&			&			\\  \hline
\end{tabular}
\footnotetext{$^1$ $g$ is the fluorescence efficiency (luminosity per molecule); $l_p$ and $l_d$ are the parent and daughter molecule scale lengths, respectively. All values are for $r_H=1$~AU, and are scaled by $r_H^{-2}$ (except for $g_{CN}$, which also varies with heliocentric velocity). Values are from \citet{A1995}.}
\end{minipage}
\end{table*}

\begin{table*}
 \centering
 \begin{minipage}{160mm}
  \caption{Measured Fluxes.}\label{tab:fluxes}
  \begin{tabular}{@{}lcccccc@{}}
  \hline
Comet	&	$r_H$ (au) &	\multicolumn{5}{c}{$F$ (erg cm$^{-2}$ s$^{-1}$)} \\ \cline{3-7}
	&	& CN (\%)& C$_3$ (\%) & C$_2$ (\%) & C$_2$ (\%)	& Continuum (\%) \\ 
      &		  & 		&	     & $\Delta v=+1$ & $\Delta v=0$ & \\ \hline
2P	&	1.9114	&	$1.43\times10^{-14}$	(2.2)	&	$1.25\times10^{-14}$	(19)	&	$<3.75\times10^{-15}$		&	$<4.49\times10^{-15}$		&	$6.81\times10^{-15}$	(46)	\\
	&	1.9221	&	$1.14\times10^{-14}$	(0.44)	&	$7.33\times10^{-15}$	(24)	&	$<2.65\times10^{-15}$		&	$<2.28\times10^{-15}$		&	$3.48\times10^{-15}$	(49)	\\
	&	1.9781	&	$9.02\times10^{-14}$	(17)	&	$<1.86\times10^{-14}$		&	$<6.29\times10^{-15}$		&	$<5.98\times10^{-15}$		&	$5.25\times10^{-15}$	(56)	\\
	&	1.9890	&	$6.58\times10^{-15}$	(50)	&	$<1.98\times10^{-14}$		&	$<3.98\times10^{-15}$		&	$<5.53\times10^{-15}$		&	$8.21\times10^{-15}$	(34)	\\
2006 K4	&	3.3713	&	$<1.33\times10^{-14}$		&	$<7.84\times10^{-15}$		&	$<5.46\times10^{-15}$		&	$<7.07\times10^{-15}$		&	$2.45\times10^{-14}$	(7.1)	\\
	&	3.3681	&	$<2.26\times10^{-14}$		&	$<1.30\times10^{-14}$		&	$<5.23\times10^{-15}$		&	$<5.49\times10^{-15}$		&	$1.86\times10^{-14}$	(9.2)	\\
	&	3.3529	&	$<9.94\times10^{-15}$		&	$<1.26\times10^{-14}$		&	$<2.63\times10^{-14}$		&	$<1.38\times10^{-14}$		&	$3.98\times10^{-14}$	(12)	\\
	&	3.3500	&	$<1.53\times10^{-14}$		&	$<1.40\times10^{-14}$		&	$<1.86\times10^{-14}$		&	$<6.84\times10^{-15}$		&	$3.03\times10^{-14}$	(11)	\\
	&	3.3472	&	$<9.02\times10^{-15}$		&	$<1.21\times10^{-14}$		&	$<5.91\times10^{-15}$		&	$<5.42\times10^{-15}$		&	$3.23\times10^{-14}$	(12)	\\
2006 OF2	&	4.7860	&	$<1.72\times10^{-14}$		&	$<2.26\times10^{-14}$		&	$<1.66\times10^{-14}$		&	$<1.10\times10^{-14}$		&	$5.30\times10^{-14}$	(3.5)	\\
	&	4.7782	&	$<1.53\times10^{-14}$		&	$<1.69\times10^{-14}$		&	$<1.40\times10^{-14}$		&	$<1.11\times10^{-14}$		&	$4.70\times10^{-14}$	(4.2)	\\
	&	4.7382	&	$<1.76\times10^{-14}$		&	$<1.82\times10^{-14}$		&	$<3.12\times10^{-14}$		&	$<2.88\times10^{-14}$		&	$8.88\times10^{-14}$	(5.9)	\\
	&	4.7304	&	$<1.50\times10^{-14}$		&	$<1.75\times10^{-14}$		&	$<2.69\times10^{-14}$		&	$<2.18\times10^{-14}$		&	$8.58\times10^{-14}$	(5.5)	\\
	&	4.7229	&	$<2.59\times10^{-14}$		&	$<1.82\times10^{-14}$		&	$<2.99\times10^{-14}$		&	$<2.05\times10^{-14}$		&	$1.10\times10^{-13}$	(4.5)	\\
2006 VZ13	&	1.0175	&	$2.58\times10^{-12}$	(10)	&	$1.74\times10^{-12}$	(53)	&	$1.69\times10^{-12}$	(15)	&	$2.85\times10^{-12}$	(29)	&	$7.40\times10^{-14}$	(11)	\\
	&	1.0165	&	$3.20\times10^{-12}$	(16)	&	$2.15\times10^{-12}$	(48)	&	$1.77\times10^{-12}$	(19)	&	$2.82\times10^{-12}$	(29)	&	$6.20\times10^{-14}$	(16)	\\
	&	1.0159	&	$4.47\times10^{-12}$	(10)	&	$2.54\times10^{-12}$	(46)	&	$3.42\times10^{-12}$	(27)	&	$6.04\times10^{-12}$	(33)	&	$9.74\times10^{-14}$	(20)	\\
	&	1.0166	&	$4.46\times10^{-12}$	(15)	&	$2.49\times10^{-12}$	(52)	&	$2.75\times10^{-12}$	(28)	&	$4.83\times10^{-12}$	(30)	&	$9.01\times10^{-14}$	(14)	\\
93P	&	2.1451	&	$7.96\times10^{-15}$	(35)	&	$<6.98\times10^{-15}$		&	$<6.50\times10^{-15}$		&	$<1.17\times10^{-15}$		&	$1.04\times10^{-14}$	(10)	\\
	&	2.1059	&	$1.69\times10^{-14}$	(74)	&	$<1.04\times10^{-14}$		&	$<5.22\times10^{-15}$		&	$<9.27\times10^{-15}$		&	$2.46\times10^{-14}$	(19)	\\
 \hline
\end{tabular}
\end{minipage}
\end{table*}

\subsection{Gas production}
The most common procedure to convert flux to a production rate is the Haser model \citep{H1957}. This model assumes the main excitation mechanism is resonance fluorescence, a spherical geometry of the coma, and an exponential decay of both the parent and daughter molecules. The fluorescence efficiencies ($g$) and scale lengths ($l_p$ and $l_d$) are scaled by $r_H^{-2}$ (where $r_H$ is the heliocentric distance) for the production rate calculations. For CN, however, there is a deviation from the $r_H^{-2}$ dependence due to the Swings effect. This is taken into account using the tables found in \citet{tatum}. The expansion velocity for all molecules is assumed to be $1~$km~s$^{-1}$, and the slit width is 2~arcsec. 

The error in the flux measurement is derived using the uncertainty of the location of the continuum. Various investigators use different values for the fluorescence efficiencies and expansion velocities, the uncertainties of which are poorly understood. Therefore, the errors in these parameters are not included in this analysis \citep[as in][]{A1995}.

For comets showing no emission in a given band, the upper limit of the production rate is determined by measuring the zero-to-peak noise in the band and multiplying by the number of pixels across the band (as in \citealt{FH1996}). For the C$_2(\Delta v=0)$ band, lines were visible for most comets; however, because of a large absorption feature within this line near 5100~\AA{}, the flux could not be measured precisely (except for 2006~VZ13). Therefore, only production rate upper limits are quoted in this case. Table~\ref{tab:Qrates} lists the calculated production rates and ratios for each comet. The numbers in parentheses give the per~cent error, which is the same per~cent error found for the flux measurement.
\begin{table*}
 \centering
 \begin{minipage}{150mm}
  \caption{Production rates and production rate ratios.}\label{tab:Qrates}
  \begin{tabular}{@{}lccccc|ccc@{}}
  \hline
Comet & r$_H$ (au) & \multicolumn{4}{c}{log Q (molecules s$^{-1}$)} &  \multicolumn{3}{c}{log Q-ratios}\\ \cline{3-6} \cline{7-9}
      &           & CN (\%)	& C$_3$ (\%) & C$_2$ (\%) & C$_2$ (\%)  &  C$_3$/CN (\%) & C$_2$/CN (\%) & C$_2$/CN (\%)\\
      &		  & 		&	     & $\Delta v=+1$ & $\Delta v=0$  & & $\Delta v=+1$ & $\Delta v=0$\\ \hline	
2P	&	1.9114	&	23.24	(2.2)	&		23.23	(19)	&	$<$	23.05		&	$<$	23.13		&		-0.011	(19)	&	$<$	-0.192		&	$<$	-0.113		\\
	&	1.9221	&	23.15	(0.44)	&		23.01	(24)	&	$<$	22.91		&	$<$	22.84		&		-0.142	(24)	&	$<$	-0.240		&	$<$	-0.305		\\
	&	1.9781	&	24.08	(17)	&	$<$	23.47		&	$<$	23.35		&	$<$	23.32		&	$<$	-0.610		&	$<$	-0.739		&	$<$	-0.760		\\
	&	1.9890	&	22.95	(50)	&	$<$	23.51		&	$<$	23.16		&	$<$	23.30		&	$<$	0.559		&	$<$	0.205		&	$<$	0.348		\\
2006 K4	&	3.3713	&	$<$	24.15	&	$<$	23.97		&	$<$	24.15		&	$<$	24.26		&	$<$	-0.179		&	$<$	0.002		&	$<$	0.114		\\
	&	3.3681	&	$<$	24.39	&	$<$	24.19		&	$<$	24.14		&	$<$	24.16		&	$<$	-0.191		&	$<$	-0.249		&	$<$	-0.228		\\
	&	3.3529	&	$<$	24.04	&	$<$	24.18		&	$<$	24.84		&	$<$	24.56		&	$<$	0.149		&	$<$	0.806		&	$<$	0.526		\\
	&	3.3500	&	$<$	24.22	&	$<$	24.23		&	$<$	24.69		&	$<$	24.26		&	$<$	0.006		&	$<$	0.467		&	$<$	0.033		\\
	&	3.3472	&	$<$	24.00	&	$<$	24.17		&	$<$	24.20		&	$<$	24.16		&	$<$	0.171		&	$<$	0.198		&	$<$	0.160		\\
2006 OF2	&	4.7860	&	$<$	25.87	&	$<$	24.90		&	$<$	25.10		&	$<$	24.92		&	$<$	-0.968		&	$<$	-0.768		&	$<$	-0.947		\\
	&	4.7782	&	$<$	25.82	&	$<$	24.78		&	$<$	25.03		&	$<$	24.93		&	$<$	-1.045		&	$<$	-0.794		&	$<$	-0.894		\\
	&	4.7382	&	$<$	25.87	&	$<$	24.79		&	$<$	25.36		&	$<$	25.33		&	$<$	-1.079		&	$<$	-0.510		&	$<$	-0.546		\\
	&	4.7304	&	$<$	25.80	&	$<$	24.77		&	$<$	25.29		&	$<$	25.20		&	$<$	-1.029		&	$<$	-0.508		&	$<$	-0.599		\\
	&	4.7229	&	$<$	26.04	&	$<$	24.79		&	$<$	25.34		&	$<$	25.17		&	$<$	-1.250		&	$<$	-0.701		&	$<$	-0.864		\\
2006 VZ13	&	1.0175	&	25.80	(10)	&		24.86	(53)	&		25.19	(15)	&		25.42	(29)	&		-0.938	(54)	&		-0.607	(18)	&		-0.380	(31)	\\
	&	1.0165	&	25.90	(16)	&		24.96	(48)	&		25.22	(19)	&		25.42	(29)	&		-0.942	(50)	&		-0.681	(25)	&		-0.480	(33)	\\
	&	1.0159	&	26.09	(10)	&		25.09	(46)	&		25.56	(27)	&		25.81	(33)	&		-1.003	(47)	&		-0.532	(28)	&		-0.285	(35)	\\
	&	1.0166	&	26.11	(15)	&		25.10	(52)	&		25.49	(28)	&		25.73	(30)	&		-1.011	(54)	&		-0.625	(32)	&		-0.380	(33)	\\
93P	&	2.1451	&	24.12	(35)	&	$<$	23.26		&	$<$	23.57		&	$<$	22.83		&	$<$	-0.862		&	$<$	-0.551		&	$<$	-1.296		\\
	&	2.1059	&	24.46	(74)	&	$<$	23.39		&	$<$	23.44		&	$<$	23.69		&	$<$	-1.070		&	$<$	-1.028		&	$<$	-0.778		\\

\hline
\end{tabular}
\end{minipage}
\end{table*}

Comet 2P shows variability in CN production, including a very large increase on 2007 August 13, as is clearly demonstrated in Fig.~\ref{fig:2P_labels}. Variabilities on the order of day are not unreasonable, as the lifetime of CN molecules is approximately $10^5$~s \citep{C1985}. On 2007 August 13, there is no large increase observed in the dust production, indicating the gas and dust production rates are not related.

When considering the comets as a group, there seems to be a linear correlation between the production rates of C$_3$, C$_2(\Delta v=0)$, and C$_2(\Delta v=1)$ with CN \citep[as reported in][]{C1987}. There does not seem to be a correlation between the production rate ratios with respect to CN and heliocentric distance, indicating gas production remains constant in a given comet over a range of distances. However, because many of the production rate values are upper limits, these statements are not conclusive.

The spectra of comet 2006~VZ13 can be compared to those in \citet{F2009} in order to determine its taxonomic class. The spectra match most closely with those of Tempel~1, and we conclude comet 2006~VZ13 is carbon-depleted.

\subsection{Dust production}
A proxy for the dust production rate can be calculated using the expression derived by \citet{A1984}:
\begin{equation}
 Af\rho=\frac{(2\Delta r_H)^2}{\rho} \frac{F_{\lambda}}{F_{\odot}}
\end{equation}
where $A$ is the Bond albedo, $f$ is the filling factor of the grains in the field of view, and $\rho$ is the circular radius of the telescope aperture at the comet (cm). $F_{\lambda}$ is the mean cometary flux in the range of $6230-6270$~\AA{}, and $F_{\odot}$ is the solar flux in the same wavelength range (both in erg cm$^{-2}$ s$^{-1}$ \AA{}$^{-1}$), calculated here using the values in \citet{A1969}. The heliocentric and geocentric distances are in au and cm, respectively. If the cometary dust is assumed to flow away from the nucleus in a uniform manner, without breakup, acceleration, or darkening, then the quantity $Af\rho$ (cm) is proportional to the dust production rate \citep{A1984,S1992}.

The continuum flux is measured in the range of 6230--6270~\AA{} of the solar-corrected spectra, where there are no known lines \citep{F1994,FH1996}. The circular radius, $\rho$, is assumed to be half the width of the rectangular slit (1~arcsec). This approximation introduces a conversion factor of $\pi$/4, a small factor which has been ignored in the dust production calculation. 

Table~\ref{tab:Afp} lists the dust production rate derived for each comet and the ratio to the production rate of CN. The per~cent uncertainty in the continuum flux is calculated using the peak-to-peak noise in the wavelength range in question. Comets with no CN line observed (2006~K4 and 2006~OF2) have lower limits of the dust-to-gas ratio listed. 

\begin{table*}
 \centering
 \begin{minipage}{60mm}
  \caption{Dust production.}\label{tab:Afp}
  \begin{tabular}{@{}lccc@{}}
  \hline
Comet	&	r$_H$ (au)	&	Af$\rho$ (cm) (\%)	& log(Af$\rho$/Q(CN)) (\%)  \\ \hline
2P	&	1.9114	&	2.02	(46)	&	-22.93	(46)	\\
	&	1.9221	&	1.06	(49)	&	-23.12	(49)	\\
	&	1.9781	&	1.84	(56)	&	-23.82	(56)	\\
	&	1.9890	&	2.96	(34)	&	-22.48	(34)	\\
2006 K4	&	3.3713	&	63.1	(7.1)	&	$>-22.35$	\\
	&	3.3681	&	48.0	(9.2)	&	$>-22.70$ 	\\
	&	3.3529	&	104	(12)	&	$>-22.02$	\\
	&	3.3500	&	79.3	(11)	&	$>-22.33$	\\
	&	3.3472	&	84.8	(12)	&	$>-22.07$	\\
2006 OF2	&	4.7860	&	398	(3.5)	&	$>-23.27$	\\
	&	4.7782	&	351	(4.2)	&	$>-23.28$	\\
	&	4.7382	&	643	(5.9)	&	$>-23.06$	\\
	&	4.7304	&	617	(5.5)	&	$>-23.01$	\\
	&	4.7229	&	788	(4.5)	&	$>-23.14$	\\
2006 VZ13	&	1.0175	&	6.76	(11)	&	-24.97	(11)	\\
	&	1.0165	&	5.80	(16)	&	-25.14	(16)	\\
	&	1.0159	&	10.3	(20)	&	-25.08	(20)	\\
	&	1.0166	&	9.77	(14)	&	-25.12	(14)	\\
93P	&	2.1451	&	6.01	(10)	&	-23.34	(10)	\\
	&	2.1059	&	12.8	(19)	&	-23.36	(19)	\\ \hline
\end{tabular}
\end{minipage}
\end{table*}

The dust-to-gas ratio stays approximately constant for each comet, except for 2P which shows a distinct decrease on 2007 August 13 (corresponding with an increase in CN production noted above); however, there seems to be an overall increase in the dust-to-gas ratio as a function of heliocentric distance. This relationship could be due to a wide range of relative abundance of refractories among the comets, or to different amounts of dust being released with the gas escaping from the surface \citep{A1995}. The overall trend is not dependent on the age or dynamical class of the comets.

\section{CONCLUSIONS}
Spectroscopic observations for five comets are reported. From these spectra, the gas production rates and production rate ratios have been calculated. There seems to be a linearity of the production rate ratios with respect to CN, agreeing with past studies \citep{C1987,A1995}. There does not seem to be a correlation between the production rate ratios and heliocentric distance. By comparing the spectra of comet 2006~VZ13 with those presented in \citet{F2009}, it is determined to be a carbon-depleted (Tempel~1 type) comet.

The dust production rate and the dust-to-gas ratio are also calculated for each comet. The ratio stays relatively constant for each comet, except for 2P due to a variability in CN production. There seems to be an overall dependence on heliocentric distance when considering the comets as a group. There is no observed dependence of the dust-to-gas ratio on the dynamical age or class of the comets.

\section*{ACKNOWLEDGEMENTS}
We would like to thank Claudio Aguilera and Alberto Miranda for acquiring the data at the CTIO. This project was supported by the Natural Science and Engineering Research Council of Canada. EU was supported by FONDECYT (Chile, grant 11060401).

\label{lastpage}


\begin{thebibliography}{}
\bibitem[\protect\citeauthoryear{A'Hearn et al.}{1984}]{A1984}
A'Hearn M., Schleicher D., Feldman P., Millis R., Thompson D., 1984, AJ, 89, 579

\bibitem[\protect\citeauthoryear{A'Hearn et al.}{1995}]{A1995}
A'Hearn M., Millis R., Schleicher D., Osip, D., Birch, P., 1995, Icarus, 118, 223

\bibitem[\protect\citeauthoryear{Arvesen, Griffin \& Pearson}{1969}]{A1969}
Arvesen J., Griffin R., Pearson B. Jr., 1969, Appl. Op., 8, 2215

\bibitem[\protect\citeauthoryear{Bus et al.}{1991}]{B1991}
Bus S., A'Hearn M., Schleicher D., Bowell E., 1991, Science, 251, 774

\bibitem[\protect\citeauthoryear{Cochran}{1985}]{C1985}
Cochran A., 1985, AJ, 90, 2609

\bibitem[\protect\citeauthoryear{Cochran}{1987}]{C1987}
Cochran A., 1987, AJ, 93, 231

\bibitem[\protect\citeauthoryear{Cochran et al.}{1992}]{C1992}
Cochran A., Barker E., Ramseyer T. Storrs, A., 1992, Icarus, 98, 151

\bibitem[\protect\citeauthoryear{Feldman, Cochran \& Combi}{2004}]{F2004}
Feldman P., Cochran A., Combi M., 2003, in Festou M., Keller H., Weaver H., eds, Comets II. Tuscon: The University of Arizona Press, p. 425

\bibitem[\protect\citeauthoryear{Fink}{1994}]{F1994}
Fink U., 1994, ApJ, 423, 461

\bibitem[\protect\citeauthoryear{Fink}{2009}]{F2009}
Fink U., 2009, Icarus, 201, 311

\bibitem[\protect\citeauthoryear{Fink \& Hicks}{1996}]{FH1996}
Fink U., Hicks M., 1996, ApJ, 459, 729

\bibitem[\protect\citeauthoryear{Haser}{1957}]{H1957}
Haser L., 1957, Bull. Soc. R. Sci. Li\`{e}ge, 43, 740

\bibitem[\protect\citeauthoryear{Newburn \& Spinrad}{1989}]{NS1989}
Newburn R., Spinrad H., 1989, AJ, 97, 552

\bibitem[\protect\citeauthoryear{Sponsetti et al.}{2006}]{S2006}
Sposetti S. et al., 2006, MPEC Circ, 2006-W03

\bibitem[\protect\citeauthoryear{Storrs et al.}{1992}]{S1992}
Storrs A., Cochran A., Barker E., 1992, Icarus, 98, 163

\bibitem[\protect\citeauthoryear{Tatum}{1984}]{tatum}
Tatum J., 1984, AA, 135, 183

\bibitem[\protect\citeauthoryear{Tody}{1993}]{T1993}
Tody D., 1993, in Hanisch R.J., Brissenden R.J.V., Barnes J., eds, Astronomical Data Analysis Software and Systems II, A.S.P. Conference Ser., Vol 52, p. 173

\end{thebibliography}
\end{document}